\documentstyle[prl,twocolumn,aps,epsfig]{revtex}        
\begin{document}
\draft
\wideabs{
\title{Dark Solitons in Bose-Einstein Condensates}
\author{S. Burger, K. Bongs, S. Dettmer, W. Ertmer, and K. Sengstock}
\address{Institut f\"ur Quantenoptik, Universit\"at Hannover,
30167 Hannover, Germany}
\author{A. Sanpera$^1$, G. V. Shlyapnikov$^{1,2,3}$, and M. Lewenstein$^1$}
\address{$^1$Institut f\"ur Theoretische Physik, Universit\"at Hannover, 30167 Hannover, Germany\\
        $^2$FOM Institute for Atomic and Molecular Physics,
        Kruislaan 407, 1098 SJ Amsterdam, The Netherlands\\
        $^3$Russian Research Center Kurchatov Institute, Kurchatov Square, 123182 Moscow, Russia}
\date{\today}
\maketitle

\begin{abstract} 
Dark solitons in cigar-shaped Bose-Einstein condensates of $^{87}$Rb are created by a
phase imprinting method.
Coherent and  dissipative dynamics of the solitons has been observed.
\end{abstract}

}
\narrowtext

The realization of Bose-Einstein 
condensation (BEC) of weakly interacting
atomic gases \cite{firstBEC}
stimulates strongly the exploration of nonlinear properties of matter waves.
This supports the new field of nonlinear atom optics, e.g., four wave mixing
in BEC's \cite{Deng1999a}, 
as well as the study of various types of excitations. 
Of particular interest are 
macroscopically excited  Bose condensed states, such as vortices and solitons.
Vortices, well known from the studies of liquid helium \cite{Donnely},
have recently been observed  
in two component gaseous condensates \cite{Matthews1999a}.

Soliton-like solutions of the Gross-Pitaevskii equation
are closely related to similar solutions in nonlinear optics describing the 
propagation of light pulses in optical fibres. 
Here, bright soliton solutions correspond to short pulses where the
dispersion of the pulse is compensated by the self-phase modulation, 
i.e., the shape of the pulse does not change. 
Similarly, optical dark solitons correspond to intensity minima 
within a broad light pulse
\cite{Kivshar1998a}.

In the case of nonlinear matter waves, 
bright solitons are only expected for an attractive interparticle 
interaction ($s$-wave scattering length $a<0$) 
\cite{Rup1995a},
whereas dark solitons, also called ``kink-states'', 
are expected to exist for repulsive interactions ($a>0$).
Recent theoretical studies discuss 
the dynamics and stability of
dark solitons 
\cite{Zhang1994aetal,Busch1998,Muryshev1999a,Fedichev1999a}
as well as concepts for their creation 
\cite{Dum1998a,Scott1998a,Dobrek1999a}.

Conceptually, solitons as particle--like objects provide a link
of BEC physics to fluid mechanics, nonlinear optics and
fundamental particle physics.


In this Letter we report on the experimental investigation of
dark solitons in cigar-shaped 
Bose-Einstein condensates in a dilute vapor of $^{87}$Rb.
Low lying excited states are produced by imprinting a local phase
onto the BEC wavefunction.
By monitoring the evolution of the density profile
we study the successive dynamics of the wavefunction.
The evolution of density minima travelling
at a smaller velocity than the speed of sound in the trapped condensate is observed.
By comparison to analytical  and numerical solutions of the 3D Gross-Pitaevskii
equation for our experimental conditions we identify these
density minima to be moving dark solitons.

In our experiment, a highly anisotropic confining potential leading to a
strongly elongated shape of the condensate 
allows us to be close
to the (quasi) 1D situation where dark solitons are expected to 
be dynamically stable
\cite{Muryshev1999a}.
Parallel to this work, soliton-like states in nearly spherical BEC's 
of $^{23}$Na are investigated at NIST \cite{Phillips1999p}.

\begin{figure}[t]
   \begin{center}
   \parbox{7cm}{
   \epsfxsize 7cm
   \epsfbox{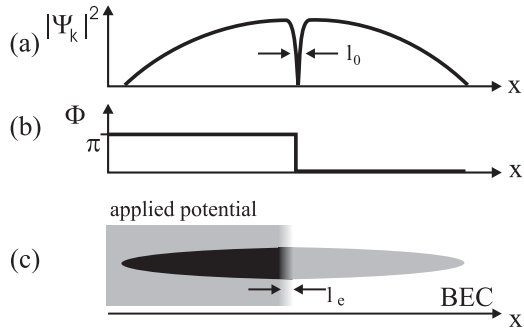}
}
   \end{center}
\caption {\small 
Density distribution (a) and phase distribution (b) of a
dark soliton state with $\Delta \phi=\pi$.
The density minimum has a width $\sim l_0$.
The scheme for the generation of dark solitons by phase imprinting 
is shown in (c), where
$l_e$ is the width of the potential edge.}        
\label{psi_bild}
\end{figure}

Dark solitons in matter waves are characterized by a
local density minimum and a sharp phase gradient of the
 wavefunction at the position
of the minimum (see Fig.\ref{psi_bild}a,b).
The shape of the soliton does not change. 
This is due to the balance between the repulsive
interparticle interaction  trying to reduce the minimum and the phase gradient trying
to enhance it.
The macroscopic wavefunction of a dark soliton in a cylindrical harmonic trap
forms a plane of minimum density (DS-plane) perpendicular
to the symmetry axis of the confining potential.
Thus, the corresponding density distribution shows a minimum at the DS-plane
with a width of the order of the (local) correlation 
length.
A dark soliton in a homogeneous
BEC of density $n_0$ is described by 
the wavefunction (see \cite{Fedichev1999a} and references therein)
\begin{equation}
\Psi_k(x) = \sqrt{n_0}\left( i\frac{v_k}{c_s}+ 
\sqrt{1-\frac{v_k^2}{c_s^2}} 
\tanh\left[\frac{x-x_k}{l_0}\sqrt{1-\frac{v_k^2}{c_s^2}}\right]\right) ,
\label{soliton_eqn}
\end{equation}
with the position $x_k$ and velocity $v_k$  
of the DS-plane, 
the correlation length $l_0=(4\pi a n_0)^{-1/2}$,
and the speed of sound $c_s=\sqrt{4\pi a n_0}\hbar/m$, where 
$m$ is the atom mass.

For $T=0$ in 1D, dark solitons are  stable. 
In this case, only solitons with zero velocity in the trap center do not
move; otherwise they oscillate along the trap axis \cite{Busch1998}.
However, in 3D at finite $T$, 
dark solitons exhibit thermodynamic and dynamical instabilities.
The interaction of the soliton with the thermal cloud causes dissipation 
which  accelerates the
soliton.
Ultimately, it reaches the speed of sound and 
disappears \cite{Fedichev1999a}.
The dynamical instability originates from the transfer of the (axial)
 soliton energy  to the radial degrees of freedom 
and leads to the undulation
of the DS-plane, and ultimately to the destruction 
 of  the soliton.
This  instability is essentially suppressed 
for solitons in cigar-shaped traps 
with a strong radial confinement \cite{Muryshev1999a}, such 
as in our experiment \cite{remark_dynsta}.

As can be seen from Eq.(\ref{soliton_eqn}), 
the local phase of the dark soliton wave
function varies only in the vicinity 
of the DS-plane, $x \approx x_k$, and is 
constant in the outer regions, with a phase difference $\Delta \phi$ between
the parts left and right to the DS-plane (see, e.g., Fig.\ref{psi_bild}b).

To generate dark solitons we apply the method of phase
imprinting \cite{Dobrek1999a}, which allows one also to create  
vortices and other  textures  in BEC's.
We apply a homogeneous potential $U_{int}$, generated by the dipole 
potential of a
far detuned laser beam, to one half of the condensate wavefunction 
(Fig.\ref{psi_bild}c).
The potential is pulsed on for a time $t_p$, 
such that the wavefunction locally acquires an 
additional phase factor 
$e^{-i\Delta\phi}$, with
$\Delta \phi = U_{int}t_p/\hbar \sim \pi$.
The pulse duration is chosen to be short compared to the
correlation time of the condensate, $t_c = \hbar /\mu$, where
$\mu$ is the chemical potential.
This ensures that the effect of the light pulse is mainly a change of the 
phase of the BEC, whereas changes of the density during
this time can be neglected.
Note, however,  that due to the imprinted phase, at larger times
one expects an adjustment of the phase and 
density distribution in the condensate. 
This will lead to the formation of a dark 
soliton and also to additional structures as 
discussed below.


In our experimental setup (see \cite{Bongs1999a}), 
condensates containing typically $1.5\times 10^5$ atoms in the
($F$=2, $m_F$=+2)-state, with less than $10\%$ of the 
atoms being in the thermal cloud, are produced every 20s. 
The fundamental frequencies of our static magnetic trap are 
$\omega_x=2\pi\times 14$Hz and $\omega_{\perp}=2\pi\times 425 $Hz
along the axial and radial directions, respectively. 
The condensates are cigar-shaped 
with the long axis ($x$-axis) oriented horizontally.

For the phase imprinting potential $U_{int}$, a blue detuned, far off resonant
laser field ($\lambda=$532nm) of 
intensity $I\approx 20$W/mm$^2$ pulsed for a time  $t_p = 20\mu$s
results in a phase shift $\Delta \phi$ of the order 
of $\pi$ \cite{remark_potential}.
Spontaneous processes can be totally neglected.
A high quality optical system is used to image an intensity profile
onto the BEC, nearly corresponding to a step function with a width of the
edge, $l_e$, smaller than 3$\mu$m (see Fig.\ref{psi_bild}c).
The corresponding potential gradient leads to a force
transferring momentum locally to the wave function and supporting the
creation of a density minimum at the position of the DS-plane for the
dark soliton.
Note that also the velocity of the soliton 
depends on $l_e$ (see Fig.\ref{velocities_bild}c).

After applying the dipole potential we let the atoms evolve within the
magnetic trap for a variable time $t_{ev}$.
We then release the BEC from the trap (switched off  within $200\mu$s) 
and take an absorption image of the density distribution
after a time-of-flight $t_{TOF}=4$ms (reducing the 
density in order to get a good signal-to-noise ratio in the images).

\begin{figure}[t]
   \begin{center}
   \parbox{9cm}{
   \epsfxsize 9cm
   \epsfbox{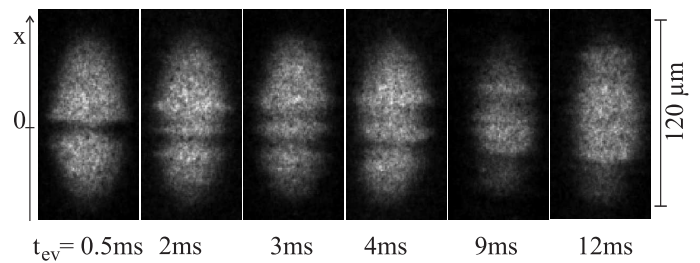}
}
   \end{center}
\caption {\small Absorption images of BEC's with kink-wise 
structures propagating in the direction of the long condensate
axis, for different evolution times in the magnetic trap, $t_{ev}$.
($\Delta \phi \sim \pi$, $N \approx 1.5\times 10^5$, $t_{TOF}=4$ms).}        
\label{absorptions_bild}
\end{figure}
\begin{figure}
   \begin{center}
   \parbox{8.5cm}{
   \epsfxsize 8.5cm
   \epsfbox{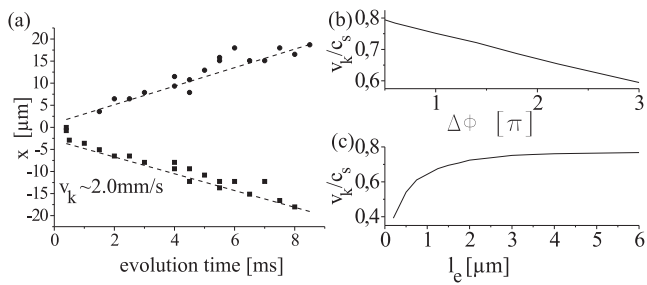}
}
   \end{center}
\caption {\small 
(a) Position of the experimentally observed density minima 
(see Fig.\ref{absorptions_bild}) 
versus evolution
time in the magnetic trap for $\Delta \phi \sim 0.5\pi$.
The dashed lines show results from the
3D simulation for $N=5\times 10^4$, $l_e=3\mu$m, 
and  $\Delta\phi = 2\pi/3$.
(b) and (c) show the dependence of $v_k/c_s$ on the
imprinted phase (for $l_e=3\mu$m) and on $l_e$ (for $\Delta \phi=\pi$)
obtained numerically in quasi 1D simulations (see [16]) 
(with $N=1.5\times 10^5$).}        
\label{velocities_bild}
\end{figure}

In series of measurements we have studied the creation and successive dynamics
of dark solitons as a function of the evolution time and the 
imprinted phase.
Fig.\ref{absorptions_bild} shows density profiles of the atomic clouds
for different evolution times in the magnetic trap, $t_{ev}$.
The potential $U_{int}$ has been applied to the part of the BEC with $x<0$.
For this measurement the potential strength 
was estimated to correspond to a phase shift of $\sim \pi$.

For short evolution times the density profile of the BEC shows a 
pronounced minimum (contrast about 40\%).
After a time of typically $t_{ev} \approx 1.5$ms a second minimum
appears. Both minima (contrast about 20\% each)
travel in opposite directions and in general
with different velocities.
Fig.\ref{velocities_bild}a) shows the evolution of these two minima 
in comparison to theoretical results obtained numerically from 
the 3D Gross-Pitaevskii equation.

One of the most important results of 
this work is that both structures
move with velocities which are smaller than the speed of sound 
($c_s\approx 3.7$mm/s for our parameters) and 
depend on the applied phase shift.
Therefore, the observed structures are different from 
sound waves in a condensate as first observed at MIT \cite{Andrews1997b}.
We identify the minimum moving slowly in the negative $x$-direction to be
the DS-plane of a dark soliton.

We have performed series of measurements with different parameter sets for $l_e$
and the product of laser intensity and imprinting time. 
The velocity of the dark soliton could thereby be
varied between $v_k=2.0$mm/s (Fig.\ref{velocities_bild}a) 
and $v_k=3.0$mm/s.
For fixed $l_e$, the velocity $v_k$ decreases with increasing $\Delta \phi$.
An increase of $\Delta \phi \sim \pi/2$ by a factor of 1.5
results in decreasing $v_k$ by 10\%, 
in agreement with theoretical
results (see Fig.\ref{velocities_bild}b,c).
For significantly reduced imprinted phase we did not observe any dark soliton
structures. For higher imprinted phase values more complex structures with several
density minima were observed.

In addition to the dark soliton, the dipole potential creates a density wave
travelling in the positive $x$-direction with a velocity close to $c_s$.
After opening the trap, a complex dynamics results in the appearance of
a second minimum behind the density wave as explained below.

To understand the generation
of dark solitons and their
behavior in the initial stages of the evolution
we have performed numerical simulations
of the 3D Gross-Pitaevskii equation.
Computing time limitations have restricted our studies to atom 
numbers below $5\times10^4$. The  simulations describe well the case $T=0$, 
but ignore the effects of thermodynamic instability.  
The latter was analyzed by using a generalization of the theory of Ref. 
\cite{Fedichev1999a}.

\begin{figure}
   \begin{center}
   \parbox{7.5cm}{
   \epsfxsize 7.5cm
   \epsfbox{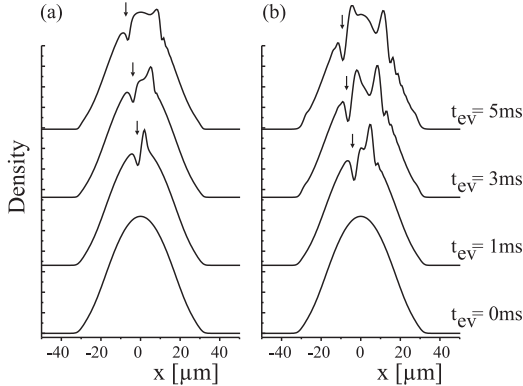}
}
   \end{center}
\caption {\small 
Evolution of the density distribution obtained numerically from the 3D simulations
for $N=10^4$ and $\Delta \phi = 2\pi /3$.
(a)
Evolution inside the magnetic trap for different
$t_{ev}$; the dark soliton is marked by an arrow.
(b)
After $t_{TOF}=4$ms, a second  density minimum is observed behind the
density wave.}        
\label{figureanna1}
\end{figure}

Our theoretical findings are summarized as follows:

\noindent 
1. 
The results of the simulations agree well with the experimental 
observations.
After applying a phase changing potential, a dark soliton moves in the 
negative $x$-direction (Fig.\ref{figureanna1}a). 
The generation of the soliton by the phase imprinting method is
accompanied by a density wave moving 
in the opposite direction. 
The maximum of the density wave travels with a velocity $\sim c_s$,
independently of the values of $\Delta \phi$ and $l_e$.
A characteristic time
for the creation of the soliton is of order $l_0/c_s$ and in our
case it does not exceed fractions of ms. 
Note that after this time the 
soliton-related phase slip in the wavefunction is affected by
a complex dynamics of the soliton generation and will be different
from the imprinted $\Delta \phi$.

\noindent
2.
For a fixed $l_e$, the increase of $\Delta \phi$ from
$\pi$ to $2\pi$, $3\pi$, \dots  
\ leads to the creation of double, 
triple solitons etc. 
BEC's with several dark 
solitons were also observed experimentally and are
currently investigated in detail.

\noindent
3.
The initial soliton velocity decreases  
as $l_e\rightarrow 0$ (Fig. \ref{velocities_bild}c).  
As observed experimentally, typical soliton velocities 
(for $l_e<3\mu$m)
are smaller than $c_s$ and grow with 
the number of atoms.
Velocities of the accompanying density waves are close to $c_s$. 
These waves move
away from the center of the trap, broaden and eventually 
vanish (Fig.\ref{figureanna1}).
This is in contrast to solitons, which are expected to oscillate in the trap,
retaining their width and absolute depth.
However,  the observation of these oscillations
would require longer lifetimes of the solitons 
(limited by dissipation to $\sim 10$ms, see below).
Within this time scale we find no signatures of dynamical 
instability and only reveal a moderate change ($<10$\%) 
of the soliton velocity, 
in agreement with our experiments.
 
\noindent 
4.
The situation changes after opening the trap and allowing 
the condensate to ballistically expand in the radial direction.
The simulation shows that the 
 soliton velocity decreases rapidly, while its width grows. 
To understand this aspect we
have used a scaling approach, similar
to that of \cite{Kagan1996a,Castin1996a},
for the radial ballistic expansion of an infinitely 
elongated condensate containing a moving kink.   
This approach  (valid for 
$\omega_{\perp}^{-1} \le t_{TOF} \le \mu/\hbar\omega_{\perp}^2$) 
predicts a soliton velocity 
$
v_k(t_{TOF}) \approx v_k(t_{ev})
\ln (2 \omega_{\perp}t_{TOF})/ \omega_{\perp}t_{TOF},
$
where $v_k(t_{ev})$ is the soliton velocity at $t_{ev}$ just before switching
 off the trap.
This result agrees very well with both  experimental data
and numerical simulations for a finite size BEC.
Moreover, in a short time $t_{TOF}\simeq \omega_{\perp}^{-1}$
 after switching off the trap the density develops a second 
 minimum located  between the density 
wave  and the dark soliton. This minimum has a width 
and depth comparable to those of the dark
 soliton (see Fig.\ref{figureanna1}b).
 Its position regarded as a function of $t_{ev}$ moves with 
a constant velocity similar to that of the soliton. 
The creation of this second minimum is a coherent phenomenon and
can be attributed to a dynamically acquired phase of 
the wave function in the 
region between the density 
wave and the dark soliton.

The results of the experiment also show clear signatures of the 
presence of dissipation originating from the interaction of the soliton with 
the thermal cloud. 
We observe a decrease of the contrast of the soliton by $\approx$50\%
on a time scale of 10ms.
This is in contradiction  with the nondissipative dynamics,  
where the contrast should even increase for the soliton moving 
away from the trap center. 
The soliton energy is then proportional 
to $n_0^{3/2}(x_k)(1- v_k^2/c_s^2(x_k))^{3/2}$ (see \cite{Fedichev1999a})
and should remain constant.
This leads to a constant absolute depth of the soliton
and hence gives the contrast proportional to  
$(1- v_k^2/c_s^2(x_k)) \sim n_0(x_k)^{-1}$. 
The decrease of the soliton contrast can therefore only be 
explained by the presence of dissipation decreasing the soliton energy.
As the life time of the soliton is sensitive to the gas temperature,
the studies of dissipative dynamics of solitons will 
offer a unique possibility 
for thermometry of BEC's in the conditions where
the thermal cloud is not discernible.

In conclusion, we have created dark solitons by a phase imprinting method and
studied their dynamics.
A detailed comparison to theory and numerical simulations allows us to identify
the creation of dark solitons
travelling with approximately 
 constant velocity smaller than the speed of sound. The initial stages of 
the evolution and the radial ballistic expansion of the sample are 
well described by the $T=0$ approach which also shows the absence of
 dynamical instabilities. The decrease of 
the soliton contrast gives a clear signature of dissipation in the soliton dynamics. 

For the study of dark solitons with even lower velocities, 
an initial density preparation of the
BEC in the magnetic trap may be useful.
This can be done, e.g., by applying adiabatically an additional blue detuned
laser beam prior to the phase imprinting pulse.
A promising attempt will be the realization of 
dark solitons in elongated
dipole traps, e.g., generated by a blue 
detuned hollow laser beam \cite{Burger1999}.
With spin as an additional degree of freedom, the dynamics 
of dark solitons in condensates containing spin-domaines or spin 
waves can be studied.

We expect that the study of soliton structures in BEC's 
opens a new direction in atomic physics, related to 
nonlinear phenomena in a dissipative environment. 

We thank D. Petrov and K. Rz\c a\.zewski for fruitful discussions
and K.-A. Suominen for help in the numerical work.
This work is supported by SFB407 of the {\it Deutsche Forschungsgemeinschaft}.
G.S.  acknowledges  support by the {\it Humboldt-Foundation},
the {\it Dutch Foundation  FOM} and the {\it Russian Foundation for Basic Studies}.


\begin{thebibliography}{10}
\bibitem{remark_dynsta}
In our case the solitons move with a velocity $\sim 0.5 c_s$ and
  are dynamically stable already at a smaller radial confinement than in the case
  of standing solitons.
\bibitem{remark_potential}
The imprinted phase is estimated from measured laser parameters and tables
of atomic data.
In addition, the onset (contrast $>10\%$)
for dark soliton creation was measured to correspond to $\Delta\phi\sim0.5\pi$,
in agreement with the theoretical results.
\end{thebibliography}



\end{document}